\documentstyle[referee]{mn}
\title{Radiative Bulk Viscosity}

\author[X. Chen and E. A. Spiegel]{Xinzhong Chen and Edward A. Spiegel\\
Astronomy Department, Columbia University, New York, NY 10027\\
eas@astro.columbia.edu\\}

\pagerange{\pageref{firstpage}--\pageref{lastpage}}
\pubyear{2000}

\begin{document}

\maketitle

\label{firstpage}

\begin{abstract}
Viscous resistance to changes in the volume of a gas arises when different
degrees of freedom have different relaxation times.   Collisions tend
to oppose the resulting departures from equilibrium and, in so doing,
generate entropy.  Even for a classical gas of hard spheres, when the
mean free paths or mean flight times of constituent particles are long,
we find a nonvanishing bulk viscosity.  Here we apply a method recently
used to uncover this result for a classical rarefied gas to radiative
transfer theory and derive an expression for the radiative stress tensor
for a gray medium with absorption and Thomson scattering.  We determine
the transport coefficients through the calculation of the comoving entropy
generation.  When scattering dominates absorption, the bulk viscosity
becomes much larger than either the shear viscosity or the thermal
conductivity.

\end{abstract}

\section{Introduction}

Both the thermal and mechanical effects of radiation on matter
play a significant role in the fluid dynamics of hot objects, from the
early universe to early stars.  The thermal effects may be reasonably
well accounted for by the use of the Eddington approximation
\cite{unn66}, though special circumstances may call for improved
descriptions.  In particular, when travel times of photons are not
negligible, retardation effects may be significant~\cite{del72}, as indeed
they are when we go to treat the dynamical effects of the radiation field.

An early description of the dynamical effects of radiation on matter
was provided by L. H. Thomas \cite{tho30} who expanded the specific
intensity in terms of the mean free path of photons, as Hilbert
had previously done for the one-particle distribution in kinetic theory
\cite{kog69}.  Each of them assumed that the relevant mean free path
is much smaller than the characteristic macroscopic length scales of the
material medium.  Though Thomas's treatment included only emission and
absorption, Masaki later extended it to include the effects of Thomson
scattering \cite{mas71,hsi76} and Compton scattering \cite{mas81}.

The effectiveness of the dissipation in those studies was characterized
by the radiative transport coefficients --- shear viscosity and
conductivity --- about whose evaluation there has been general agreement
since Thomas's derivation of the stress tensor for a matter-radiation
mixture with short photon mean free paths~\cite{wei72,mih84}.
But in recent years, particular attention has been given to the bulk
viscous effects, which are of interest for applications to cosmology and,
for that case, different expressions have been proposed.

A simple relativistic gas has bulk viscosity~\cite{ste72} because the
ultrarelativistic particles and particles with small energies respond to
volume changes at different rates and so are driven out of equilibrium
with each other under expansion.  Analogously, in radiative fluid
dynamics, the temperatures of the matter and the radiation may not be
equal in nonequilibrium processes.  This too can be a cause of a bulk
viscosity~\cite{wei71,mih84}, even though photons are ultrarelativistic.
Weinberg gave an expression for the radiative bulk
viscosity in the Thomas case with only emission and absorption
\cite{wei71}.  Subsequent work~\cite{and77,ste72}, also for media with
only emission and absorption but based on other methods, gave different
expressions for the forms of the transport coefficients.  However, those
results have been shown to be equivalent to Weinberg's in the first order
in photon mean free path.

Additional studies have included scattering but, in those more complicated
calculations, the bulk viscosity has either not been derived or not
extracted explicitly~\cite{mas71,hsi76,str76,mas81,tho81}.  Hence our aim
here is to determine the bulk viscosity coefficient from a calculation of
the entropy generation in the comoving frame of the matter, once we have
obtained suitable approximations for the radiative stress tensor.  Since
the previous studies have been limited to cases where the mean free
paths of photons are much smaller than the prevalent characteristic
scales, the problem of treating situations where transparent regions
form in the medium has been left open.  This issue must be confronted at
the edges stars and disks, in intense turbulence and in photon bubbles.  A
case of particular interest arises in models of the early universe when
the mean flight times of photons are comparable to the age of the
universe.

The difficulties posed by long mean free paths are less severe for thermal
than for dynamical problems since, as we mentioned, the radiative
smoothing of temperature fluctuations is reasonably well described by the
Eddington approximation, both in optically thick and thin regions.  A
reason for this is that the Eddington approximation represents a
summation of contributions from terms of all orders in a mean free path
expansion \cite{unn66}.  However, since the Eddington approximation
leads to a diagonal stress tensor, it fails completely to describe shear
viscosity~\cite{and72}.  Moreover, it does not give a good representation
of the bulk viscosity that opposes changes in volume.

Methods based on truncation of the moment hierarchy at higher moments than
the pressure tensor do lead to shear viscosity~\cite{tho81,str97}, but
they give complicated equations and are difficult to use.  Another
approach is to use a more general resummation procedure than that leading
to the Eddington approximation, one that provides a nonlocal viscous shear
tensor.  Though this has been done for the case of pure absorption
and emission~\cite{cherad}, it has not been carried out with scattering
included.  Nor is this issue to be considered here. Our purpose in this
paper is rather to confront the problem of bulk viscosity for a medium
where the photon mean free path is not short and where we also have to
deal with scattering processes.  For this, we shall adopt a procedure
that has recently been introduced in classical kinetic theory \cite{bgk1}
to generalize the usual Navier-Stokes equations.  The generalized
equations exhibit resistance to volume changes when the particle mean free
paths are long, even for a gas of hard spheres.

To test the results obtained with the new method, we have calculated the
thicknesses of shock waves and the propagation speeds of ultrasonic
waves for classical gases.  The results found are in good agreement with
experiment, whereas those obtained from the Navier-Stokes equations do not
fare well.  Though the limits of validity of the new approach are still
being probed, the results seem good enough to make the application to the
problems of radiative fluid dynamics seem worthwhile.  Since the methods
that have been used heretofore to compute the radiative dissipation terms
parallel the Chapman-Enskog procedure \cite{kog69,cer88}, we may expect to
encounter differences from previous results in this case as well.

In what follows, we describe an expansion in photon mean free path for
solutions of the transfer equation, including the effects of Thomson
scattering, and develop a stress tensor for use in media whose photon
mean free paths need not be small.   From this, we compute the rate
of comoving entropy generation and provide a formula for bulk viscosity.
In a later paper, we shall use this result to estimate the entropy
generation in the standard model of cosmology.

\section{Transfer Theory}

\subsection{The radiation field}

We consider radiative transfer in a fluid medium with density $\rho$ and
velocity $u^\mu$, each
depending on location in space-time, $x^\mu$, where $\mu=0,1,2,3$.
There are two reference frames of basic interest in this work, an
inertial frame, or system frame, and the comoving frame of the matter.
In the system frame, we have \begin{equation} u^\mu = \gamma(1, {\bf
v}), \qquad \qquad \gamma = (1-{\bf v})^{-1/2}
\label{fourv} \end{equation}
while in the comoving frame $u^\mu=(1,{\bf 0)}$.  Here, as in the
following, we assume units in which the speed of light {\it in vacuo}
and Planck's constant are each unity.  We also adopt the signature
in which the Minkowski metric is diag$(1, -1, -1, -1)$.

The four-momentum, $p^\mu$, of a photon satisfies $p_\mu p^\mu=0$, so
that we may define a basic null vector, \begin{equation}
n^\mu = p^\mu/\nu = (1,{\bf n}), \label{nmu} \end{equation}
that characterizes the direction of the photon's motion in spacetime.
With $\nu$ as the
frequency in the inertial frame, the rest frequency (as seen comoving
locally with the medium) is given by the Doppler formula
\begin{equation}
\tilde \nu = u_\mu p^\mu = \gamma\nu(1-{\bf v}\cdot{\bf n}) \ .
 \label{tldenu} \end{equation}

The coordinates of the phase space of photons are $x^\mu$ and $p^\mu$,
which we shall sometimes denote as $x$ and $p$ for brevity.  We shall
describe the density of photons in phase space by the one-particle
distribution function, $f(p, x)$.  This quantity is a scalar and its
integral over the invariant volume in momentum space gives the photon
number density in space.  To perform this integration, we introduce the
invariant volume in phase space, $dP = \nu d\nu d\Omega$, where
$d\Omega$ is the element of solid angle~\cite{lan84}.

A principal quantity of interest in this work is the stress-energy
tensor of the radiation field, \begin{equation} T^{\mu\nu} = \int
p^\mu p^\nu f\, dP = \int \int n^\mu n^\nu {\cal I} \; d\Omega d\nu \
, \label{tmunu} \end{equation} where the last term on the right
introduces the specific intensity, defined as
\begin{equation}
{\cal I} = \nu^3 f \ . \label{spI} \end{equation}

A revealing way of writing the stress tensor in terms of basic
moments of the radiation field is obtained by defining the
directional vector with respect to the moving matter, \begin{equation}
l^\mu = n^\mu - u^\mu\ ,  \label{defl} \end{equation}
so that \begin{equation}
\l_\mu l^\mu = -1 \qquad {\rm and} \qquad \l_\mu u^\mu = 0 \ .
\label{prop}
\end{equation}
Hence $l^\mu=(0,{\bf l})$, where ${\bf l}$ is a unit three-vector in
the spatial direction of photon motion in the comoving frame.
The angular moments of the radiation field are then
\begin{equation}
{\cal E} = \int {\cal I} d\Omega\ , \qquad {\cal F}^\mu = \int {\cal
I} l^\mu d\Omega \ , \qquad {\cal P}^{\mu\nu} = \int {\cal I} l^\mu
l^\nu d\Omega \ ,
\label{chrommo} \end{equation}
and their frequency-integrated forms are
\begin{equation} E=\int_0^\infty {\cal E}  d\nu \qquad
F^\mu=\int_0^\infty {\cal F}^\mu d\nu \qquad P^{\mu\nu}=\int_0^\infty
{\cal P}^{\mu\nu} d\nu \ . \label{freq}
\end{equation}
These moments together make up the energy-momentum stress tensor which
may now be written as \begin{equation} T^{\mu\nu}=E u^\mu u^\nu +F^\mu
u^\nu +F^\nu u^\mu +P^{\mu\nu} \label{tmunu1} \ .
\end{equation}

\subsection{The Transfer equation}

When both absorption and scattering occur, the transfer equation
takes the general form~\cite{sim63} \begin{equation}
p^{\mu}\partial_{\mu}f=\rho(\alpha-\beta f)+ \rho\int \Re (p,
p')f(p')dP'-\rho\int \Re (p', p)f(p)dP'
\label{eq1}
\end{equation}
where $\alpha$ and $\beta$ are scalars characterizing absorption and
emission and the kernel $\Re(p',p)$ is the differential
cross-section for scattering a photon from four-momentum $p'$ into
$p$.  For brevity, we have not indicated the dependence
of $f$ on $x$ in the transfer equation and have omitted the effects
of gravity.  We here consider only the case of Thomson scattering
for which we have~\cite{mas71,tuc75}
\begin{equation} \Re (p, p')=\frac{3}{4}[1+({\bf l}\cdot {\bf l}')^2]
\, \delta(\tilde\nu-\tilde\nu')\frac{\sigma}{4\pi}
\label{eq2}
\end{equation}
where $\sigma$ is the Thomson cross-section.

We may factor $f$ out of the second integral of (\ref{eq1}) and note that
with $i=1,2,3$ we have \begin{equation} \int l^i l^j \, d\Omega = -
{4\pi\over 3} \delta^{ij} \label{delt}
\end{equation}
where $\delta^{ij}$ is the Kronecker symbol.  Then the last term
in (\ref{eq1}) reduces to $-\rho\sigma f$.  Since $f$ is a scalar, we may
use (\ref{spI}) to find the transformation law for the specific intensity.
Also, we may compare the covariant form of the transfer equation with the
usual one for the comoving frame, as Thomas~\cite{tho30} did, to find
$\beta = \tilde\nu \kappa(\tilde\nu)$, where $\kappa$ is the absorption
coefficient, and $\alpha = \tilde\nu^{-2}j(\tilde\nu)$, where $j$ is
the emissivity.  Since $\alpha$, $\beta$ and $\tilde\nu$ are scalars,
we then know the transformation rules for $j$ and $\kappa$ once we
introduce the (relativistic) Doppler formula (\ref{tldenu}).

The equation of transfer may now be written as~\cite{hsi76}
\begin{equation}
\varepsilon n^\rho {\cal I}_{,\rho} = \hat \kappa {\cal S} - {\cal I}
+ {3\hat \sigma \over 16 \pi}\left({\cal E} + l_\rho l_\sigma {\cal
P}^{\rho\sigma}\right) \ ,
\label{retrans} \end{equation}
where \begin{equation} {\cal S} = \nu^3 \alpha/\beta \ , \label{sou}
\end{equation}
\begin{equation}
\varepsilon = [\rho(\kappa+\sigma)]^{-1} \ , \label{varep}
\end{equation}
\begin{equation}
\hat \kappa=\frac{\kappa}{\kappa+\sigma} \ , \qquad \qquad \hat
\sigma=\frac{\sigma}{\kappa+\sigma} \label{hat} \end{equation} and a
comma with subscript $\mu$ indicates differentiation with respect to
$x^\mu$.  We have replaced $\delta_{ij}{\cal P}^{ij}$ by
$\delta_{\rho\sigma}{\cal P}^{\rho\sigma}$, to which it is equal, for
cosmetic reasons.

Expressions for the moments that appear in the transfer equation may
be computed from the equation itself.  In doing this, we shall assume
that the medium is grey, although we could save appearances by
introducing suitable mean absorption coefficients.  We shall not
follow that practice here since it leads to serious
complications in moving media, such as tensorial absorption
coefficients, and has not repaid the effort involved.

Next we derive expressions for the monochromatic moments defined
in (\ref{chrommo}) by taking suitable moments of (\ref{retrans}).
For this purpose, we define \begin{equation}
h^{\rho\sigma} = -{3\over 4\pi} \int l^{\rho} l^\sigma d\Omega \; ,
\label{hrs} \end{equation}
which, in terms of the basic tensors of the problem, may be expressed as
\begin{equation}
h^{\rho\sigma} = \eta^{\rho\sigma} - u^\rho u^{\sigma} \ . \label{h}
\end{equation}
Similarly, we have \begin{equation} \int l^\mu l^\nu l_\rho
l_\sigma \, d\Omega = {4\pi\over 15} (h^{\mu\nu}h_{\rho\sigma} +
h^{\mu}_{\rho} h^{\nu}_{\sigma} + h^\mu_\sigma h^\nu_\rho) \
. \label{4h} \end{equation}

Then, when we integrate (\ref{retrans}) over all solid angle and
note that $h_{\mu\nu}{\cal P}^{\mu\nu}=-{\cal E}$, we find that
\begin{equation}
{\cal E} = 4\pi {\cal S} - {\varepsilon\over \hat \kappa} \int n^\rho
{\cal I}_{,\, \rho}\, d\Omega \ . \label{calE1} \end{equation}
Further, on applying the second definition in (\ref{chrommo}) to
(\ref{retrans}), we get \begin{equation} {\cal F}^\mu=-\varepsilon\int
l^\mu n^\rho{\cal I}_{,\rho}\,d\Omega\, \ .
\label{calF}\end{equation}
Finally, on multiplying (\ref{retrans}) by $l^\mu l^{\nu}$ and
integrating over angle we find that \begin{equation} {\cal
P}^{\mu\nu} = -{40\pi\hat
\kappa\over 3(10-\hat \sigma)} {\cal S} h^{\mu\nu} - {3\hat
\sigma\over 10 - \hat \sigma}{\cal E} h^{\mu\nu} -{10 \varepsilon\over
10-\hat \sigma} \int l^\mu l^\nu n^\rho\, {\cal I}_{,\rho}\, d\Omega \; .
\label{calP} \end{equation}

With the moments expressed in this way, we obtain the expression
\begin{equation} {\cal E} + l_\rho l_\sigma {\cal P}^{\rho\sigma} =
{16\pi\over 3} {\cal S} - {2\varepsilon(5+\hat\sigma)\over \hat \kappa
(10 - \hat \sigma)} \int n^\rho\, {\cal I}_{,\rho}\, d\Omega -{10
\varepsilon\over 10-\hat \sigma}l_\mu l_\nu \int l^\mu l^\nu n^\rho\,
{\cal I}_{,\rho}\, d\Omega \ .
\label{PE} \end{equation}
Hence we may write the transfer equation in the compact form
\begin{equation}
{\cal I} = {\cal S}
- \varepsilon {\cal L} {\cal I} \label{reret}
\end{equation} where we have introduced the linear operator
\begin{equation} {\cal L} =n^\rho\partial_\rho + c_1\int
d\Omega\,n^\rho\, \partial_\rho+c_2 l_\mu l_\nu \int d\Omega\,l^\mu
l^\nu n^\rho\, \partial_\rho \label{calL}
\end{equation} with $\partial_\rho = \partial/\partial\, x^\rho$,
\begin{equation}c_1={3\hat \sigma(5+\hat\sigma)\over 8\pi\hat
\kappa (10 - \hat \sigma)}\label{c1}\end{equation} and
\begin{equation}c_2={15\hat \sigma\over 8\pi(10-\hat \sigma)} \ .
\label{c2}\end{equation}
When there is no scattering, $\hat \sigma=0$ and the foregoing
equations reduce to those for the case of pure absorption.

\section{The Radiative Pressure Tensor}

\subsection{The Expansion Procedure}

We expand the intensity ${\cal I}$ as \begin{equation} {\cal
I}=\sum_{m=0}^{\infty}{\cal I}_{(m)}\varepsilon^m \ .
\label{exp}
\end{equation}
From ${\cal I}_{(m)}$, we can calculate the coefficients of
the expansions of the monochromatic moments and their
frequency-integrated counterparts from the definitions
\begin{equation}
{\cal E}_{(m)} = \int {\cal I}_{(m)} d\Omega\ , \qquad {\cal
F}^\mu_{(m)} = \int {\cal I}_{(m)} l^\mu d\Omega \ , \qquad {\cal
P}^{\mu\nu}_{(m)} = \int {\cal I}_{(m)} l^\mu l^\nu d\Omega
\label{chrommom} \end{equation}
and
\begin{equation} E_{(m)}=\int_0^\infty {\cal E}_{(m)}  d\nu \qquad
F^\mu_{(m)}=\int_0^\infty {\cal F}_{(m)}^\mu d\nu \qquad
P^{\mu\nu}_{(m)}=\int_0^\infty {\cal P}^{\mu\nu}_{(m)} d\nu \
. \label{freqm}
\end{equation}

When we introduce the expansion (\ref{exp}) into the transfer equation
(\ref{reret}) and demand that the expanded equation is satisfied term
by term, we get \begin{equation} {\cal I}_{(0)} ={\cal S} \, ,
\label{f0}
\end{equation}
as in comoving local thermodynamic equilibrium.  For the higher orders
with $m\geq 1$, we obtain \begin{equation} {\cal I}_{(m)}=-{\cal L}
{\cal I}_{(m-1)} \label{Im}
\end{equation}
where ${\cal L}$ is defined in (\ref{calL}).

The problem studied is really one of a mixture of radiation and matter
and a more complete description would begin with coupled transport
equations.  In the present version, we simply proceed as if
the material properties are given, as did Thomas~\cite{tho30}.
In doing this, we consider that the properties of the medium
are expressed, not as functions of $x^\mu$, but in terms of the
basic fields, $T$, $\tilde \nu$, $n^\rho$.  Therefore, in the simplest
case, where we operate on a function of $T$, $\tilde \nu$ and $n^\rho$
only, we may write \begin{equation}
\partial_\mu =
T_{,\mu}\,\partial_T+\tilde\nu_{,\mu}\,\partial_{\tilde\nu}
+n^{\sigma}_{,\mu} \partial_{n^\sigma} \ .  \label{partialmu0}
\end{equation}

On recalling that $\tilde\nu =u_\mu p^\mu$ and that $p^\mu$ does
not depend on coordinate, we obtain \begin{equation}
\tilde{\nu}_{,\mu} = p_\nu u^\nu_{, \mu} \ . \label{numu}
\end{equation} Similarly, on differentiating
$p^{\mu}=n^{\mu}\tilde\nu$ and making use of (\ref{numu}) and the
identity $u^\mu u_{\mu,\rho}=0$, we get \begin{equation} n^\mu_{\,
,\nu}=-n^\mu n^\rho u_{\rho,\nu} \ . \label{ncomma}
\end{equation}
So (\ref{partialmu0}) takes the form \begin{equation} \partial_\mu =
T_{,\mu}\partial_T\,+\tilde\nu n^\rho
u_{\rho,\mu}\,\partial_{\tilde\nu} \ -n^\sigma n^\rho
u_{\rho,\mu}\,\partial_{n^\sigma} \; .
\label{partialmu2} \end{equation}
If we further assume that, like the Planck function, ${\cal S}$ depends on
$\tilde \nu$ and $T$ only through the combination $\tilde \nu/T$, we
deduce from (\ref{partialmu2}) that \begin{equation}
{\cal S}_{,\mu} = \left(T_{,\mu}\, - T n^\rho
u_{\rho,\mu}\right) \partial_T {\cal S}\; . \label{smu} \end{equation}

We note that in the higher orders some of the derived quantities will
depend explicitly on the derivatives of $T$ and $u^\rho$ as well as on the
fields themselves.  Thus, more generally, when we go to higher order
developments, as we did in computing the shear viscosity~\cite{cherad}, we
must allow for a dependence on those derivatives.  However, here we go
only to first order in photon mean free path and that complication does
not arise; (\ref{partialmu2}) is therefore sufficient for our purposes.

Returning to the calculation of ${\cal E}$, we see from
(\ref{calE1}) that ${\cal E}_{(0)}=4\pi {\cal S}$ and that, for
$m\geq 1$,
\begin{equation}
{\cal E}_{(m)} = -
{1\over \hat \kappa} \int n^\rho\, \partial_{\rho}
{\cal I}_{(m-1)}\, d\Omega \ .  \label{calEm} \end{equation}
Also, from (\ref{calF}) we have ${\cal F}^{\mu}_{(0)}=0$ and, for
$m\geq 1$, \begin{equation}
{\cal F}^{\mu}_{(m)} = -\int l^\mu n^\rho\, \partial_\rho
{\cal I}_{(m-1)}\,d\Omega\, .  \label{calFm}\end{equation}
Then from (\ref{calP}) we find ${\cal P}_{(0)}^{\mu\nu}=-\frac{4\pi}{3}
{\cal S}h^{\mu\nu} $ and, for $m\geq 1$,
\begin{equation}
{\cal P}^{\mu\nu}_{(m)} = -{3\hat \sigma\over 10 - \hat \sigma} {\cal
E}_{(m)} h^{\mu\nu} -{10\over 10-\hat \sigma} \int l^\mu l^\nu
n^\rho\, \partial_{\rho}\,{\cal I}_{(m-1)}\, d\Omega \ .  \label{calPm}
\end{equation}

\subsection{The First-Order Stress Tensor}

From the recursion relation (\ref{Im}) we have
\begin{equation}
{\cal I}_{(1)}={\cal L} {\cal I}_{(0)}= {\cal L} {\cal S} \; .
\label{deriv} \end{equation}
Then from (\ref{smu}) and (\ref{calEm}) we get
\begin{equation}
{\cal E}_{(1)}=-\frac{4\pi}{\hat \kappa}({1\over 3} T \theta +\dot
T) {\cal S}_{,T} \end{equation} where \begin{equation}
\theta=u^\mu_{\ ,\mu} \label{thet} \end{equation} and $\dot
T=u^{\mu} T_{,\mu}$.  On integrating over frequency, we find
\begin{equation}
E_{(1)}=-\frac{4\pi}{\hat \kappa}(\frac{4}{3}S\theta+\dot{S}),
\label{sce1}
\end{equation}
where \begin{equation} S=\int_0^\infty {\cal S} d\nu = aT^4 \label{T4}
\end{equation} and $\dot S = u^\mu S_{,\mu}$.

In a similar way, we find for the frequency-integrated first-order
radiative flux,
\begin{equation}
F_{(1)}^{\mu}= \frac{4\pi}{3}[S_{,\rho}-4S\dot{u}_\rho]h^{\mu\rho},
\label{scf1}
\end{equation}
while for the frequency-integrated pressure tensor in first order, we
obtain
\begin{equation}
P_{(1)}^{\mu\nu} = \frac{40\pi}{10-\hat \sigma} \left[-\frac{3\hat
\sigma}{40\pi}h^{\mu\nu}
E_{(1)}+\frac{h^{\mu\nu}}{3}\dot{S}+\frac{4S}{15}
(\tau^{\mu\nu\rho\sigma}u_{\rho,\sigma}+\frac{5}{3} \label{mes}
h^{\mu\nu}\theta)\right] \end{equation} where \begin{equation}
\tau^{\mu\nu\rho\sigma}=h^{\mu\rho}h^{\nu\sigma}+
h^{\mu\sigma}h^{\nu\rho}-\frac{2}{3}h^{\mu\nu}h^{\rho\sigma}\ .
\end{equation}
We may then introduce (\ref{sce1}) and so write the pressure tensor to
this order as \begin{equation}
P^{\mu\nu} =-\frac{4\pi}{3}h^{\mu\nu}\left[S-\frac{\varepsilon}{\hat
\kappa} (\frac{4}{3}S\theta+\dot{S})\right]+\Xi^{\mu\nu}\label{close}
\end{equation}
where \begin{equation}
\Xi^{\mu\nu} = \mu\, \tau^{\mu\nu\rho\sigma}u_{\rho,\sigma}
\label{Xi}\end{equation}
is the shear tensor and the shear viscosity coefficient is that found
by Thomas and Masaki, \begin{equation}
\mu=\frac{32\pi S}{3\rho (10 \kappa + 9 \sigma)} \ . \label{visc}
\end{equation}
As mentioned, we have elsewhere computed a nonlocal expression
for the shear tensor for the case of pure absorption~\cite{cherad} and
expect that a similar generalization may apply in the case with
scattering, though we have not yet carried out the necessary calculations
to confirm this.

From (\ref{scf1}), we may identify the thermal conductivity
as the coefficient of $T_{,\rho}$:
\begin{equation} \chi=\frac{16\pi S}{3T\rho(\kappa+\sigma)} \ .
\end{equation}
The second term in the flux, with coefficient $\xi=T\chi$, has been
considered to be a purely relativistic effect \cite{mih83}, but there is a
classical counterpart when the particles have long mean free
paths~\cite{bgk1}.  We shall see in the next section how these transport
coefficients, as well as the bulk viscosity, can be deduced from the
entropy generation rate.

\section{Entropy Generation and Bulk Viscosity}

To clarify the roles of the various terms in the radiative stress
tensor, we compute their contributions to the entropy generation.
In leading order, this is related to the deviation
of the stress tensor from its equilibrium value, which we
approximate to first order in $\varepsilon$ as \begin{equation}
\Delta T^{\mu\nu} = T^{\mu\nu} - T^{\mu\nu}_{(0)} = \varepsilon
T^{\mu\nu}_{(1)} \label{deltaT} \ ,  \end{equation}
where \begin{equation}
T^{\mu\nu}_{(1)}=E_{(1)} u^\mu u^\nu +F_{(1)}^\mu u^\nu + F_{(1)}^\nu
u^\mu +P_{(1)}^{\mu\nu}\ . \label{T(1)} \end{equation}

The rate of entropy production is the four-divergence of the entropy
flux, $\Sigma^{\mu}$, whose expression as given by Weinberg~\cite{wei72}
and recalled in Appendix A is \begin{equation}
T\Sigma^\mu_{,\mu}=\Delta T^{\mu\nu}\left[u_{\mu,\nu}- u_\mu (\log
T)_{,\nu}\right].  \label{n6} \end{equation}
For the components of $T^{\mu\nu}_{(1)}$ we introduce (\ref{sce1}),
(\ref{scf1}) and (\ref{mes}) and obtain, after some simple
rearrangements, \begin{equation}
\rho(\kappa+\sigma)T\Sigma^\mu_{,\mu}=-E_{(1)}\left[u^\nu
(\log T)_{,\nu} +\frac{1}{3}\theta\right] + \Xi^{\mu\nu} u_{\mu,\nu}
+F_{(1)}^\mu\left[u^\nu u_{\mu,\nu}- (\log T)_{,\mu}\right] .
\label{n12}
\end{equation}

Since $S=aT^4$, we can rewrite (\ref{sce1}) as
\begin{equation}
E_{(1)}=-\frac{16\pi S}{\hat \kappa}\left[ u^\rho (\log
T)_{,\rho}+\frac{1}{3}\theta\right],
\label{n4.1}
\end{equation}
which we use in the first term of (\ref{n12}).  To simplify the second
term in (\ref{n12}), we note that, on account of the form of
$\Xi^{\mu\nu}$, we encounter the expression
$u_{\mu,\nu}\tau^{\mu\nu\rho\sigma}u_{\rho,\sigma}$.  We may use the
identity $h_{\mu\nu}\tau^{\mu\nu\rho\sigma}=0$ to show that
\begin{equation}
u_{\mu,\nu}\tau^{\mu\nu\rho\sigma}u_{\rho,\sigma}
=\frac{1}{2}(u_{\mu,\nu}+u_{\nu,\mu}-\frac{2}{3}h_{\mu\nu}\theta)
\tau^{\mu\nu\rho\sigma} u_{\rho,\sigma}\, .
\label{n15.5}
\end{equation}
Then we find that
\begin{equation}
u_{\mu,\nu}\tau^{\mu\nu\rho\sigma}u_{\rho,\sigma}
=\frac{1}{4}(u_{\mu,\nu}+u_{\nu,\mu}-\frac{2}{3}h_{\mu\nu}\theta)
\tau^{\mu\nu\rho\sigma} (u_{\rho,\sigma}+
u_{\sigma,\rho}-\frac{2}{3}h_{\rho\sigma}\theta)\; .
\label{n15}
\end{equation}

All this adds up to \begin{eqnarray}
\Sigma^\mu_{,\mu}
& = & -\eta h^{\mu\nu} \left(\frac{T_{,\mu}}{T}-u^\rho
u_{\mu,\rho}\right)\left(\frac{T_{,\nu}}{T} -u^\rho
u_{\nu,\rho}\right) +\zeta\theta^2 \nonumber \\ & + & \frac{2\mu}{T}
h^{\mu\rho}h^{\nu\sigma}
(u_{\mu,\nu}+u_{\nu,\mu}-\frac{2}{3}h_{\mu\nu}
\theta)(u_{\rho,\sigma}+
u_{\sigma,\rho}-\frac{2}{3}h_{\rho\sigma}\theta) \label{n16}
\end{eqnarray}
where
\begin{equation}
\chi=\frac{16\pi S}{3T\rho(\kappa+\sigma)} \; , \label{n19}
\end{equation}
\begin{equation}
\mu=\frac{32\pi S}{3(10 \kappa + 9 \sigma)}\label{n18}
\end{equation}
and
\begin{equation}
\zeta =\frac{16\pi S}{\rho \kappa T} \left[\frac{1}{3} +
\frac{u^\alpha T_{,\alpha}}{T\theta}\right]^2  \ .
\label{xi0}
\end{equation}

To see the meaning of these results, we note that for any four-vector
$v^{\mu}$, whose components in the comoving frame are $(v^0,v)$, we
have \begin{equation}
h^{\mu\nu}v_{\mu}v_{\nu}=\delta^{ij}v_iv_j=-\parallel\vec{v}
\parallel^2\leq 0. \end{equation}
Similarly, for any rank-2 tensor $w^{\mu\nu}$, we can show that
$h^{\mu\rho}h^{\nu\sigma}w_{\mu\nu}w_{\rho\sigma}\geq 0$.  So we find
that the rate of entropy generation is never negative as long as all these
coefficients are positive.

The first term in (\ref{n16}) is the entropy generation produced by the
drag of the radiative flux, which is an effect of radiative thermal
conductivity, with due allowance for the acceleration of the matter.  The
second term represents the entropy generation produced by volume changes
and the coefficient in this term, $\zeta$, is the radiative bulk
viscosity.  Finally, the third term is a rate of entropy increase caused
by shear, with the coefficient of radiative shear viscosity, $\mu$.

We see from (\ref{xi0}) that the bulk viscosity depends on $u^\rho
T_{,\rho}= \dot T$.  In depending on a rate, our bulk viscosity differs
qualitatively from that derived by Weinberg \cite{wei72} who found
a bulk viscosity of the form \begin{equation}
\zeta_W=\frac{16\pi S}{\rho \kappa T}\left[\frac{1}{3}-\left(
\frac{\partial P}{\partial e}\right)_n\right]^2 \; .
\label{xi0wein}
\end{equation}
For a radiation-dominated situation, in which
$\left(\frac{\partial P}{\partial e}\right)_n\approx \frac{1}{3}$,
we get  $\zeta_W\approx 0$.

The principal reason for the difference between the results (\ref{xi0})
and (\ref{xi0wein}) is that Weinberg followed the standard practice in
transport theory of using a lower order condition (here
$T_{(0),\nu}^{\mu\nu}=0$) to eliminate the time derivative from the
pressure tensor \cite{wei72,ber88}.  Thus he obtained
$\dot{S}=-4S\left(\frac{\partial p}{\partial e}\right)_n\theta$ (see
Appendix B) and was led to formula (\ref{xi0wein}) for the bulk viscosity.
While this does simplify the derivation of (\ref{xi0wein}), it also
shrinks its domain of validity.  In avoiding the use of this condition
in our derivation, we obtain a result that is capable of describing
relatively rapid processes and narrow structures.  Moreover, we find that
the bulk viscous pressure can be large even for the radiation-dominated
case.

It is interesting that with both approaches, as the absorption
coefficient gets smaller, the bulk viscosity grows larger.  This is not
so surprising since, for large absorption coefficient, the temperature
difference between radiation and matter is small.  Still, the tendency
toward large bulk viscosity at small absorption coefficient is noteworthy
and is reminiscent of the sharply different behaviors of fluids with
zero viscosity and those with viscosity tending to zero.  Of course, since
the effect of Compton scattering will be like that of absorption, the
limit of zero absorption strictly does not apply to the general case.
Nevertheless, we may expect to find effective heating of expanding media
in some situations.

\section{Conclusion}

From its very beginnings, the subject of radiative viscosity has been
vexatious, as several of the founders of modern radiative transfer
theory --- Milne, Eddington, Jeans --- found in their attempts to
compute the shear viscosity \cite{mih83}.  It was not until
Thomas \cite{tho30} did the problem relativistically that agreement
was reached on the value of the shear viscosity, to leading order.
What is interesting is that the agreement seems to have been reached
on purely theoretical grounds since there do not appear to be direct
measurements confirming Thomas's results.

In the matter of bulk viscosity, the situation is even more troubling
in that the calculations are subtler and there are by now many
contributors to the growing literature on this issue.  It is
conceivable that disagreements over bulk viscosity may also be settled
by common consent, but we feel this is unlikely.  As we have noted,
in previous work, the simplification of introducing equilibrium
approximations into the lower order theory is used, in the spirit of the
Chapman-Enskog procedure \cite{kog69,cer88}.  For gases of classical
particles, this approximation leads to the conclusion that a simple
classical gas does not have a bulk viscosity.  However, when this
restrictive approximation is not made, we have found \cite{bgk1}, on
using the relaxation (or BGKW) model \cite{kog69,cer88}, that a classical
rarefied gas may have an effective bulk viscosity and that the gas obeys
macroscopic equations that differ from the Navier-Stokes equations when
the particle mean free paths become long. In the classical case, one has
recourse to experiments to test the conclusions and these have lent to
support to our conclusion.  In the radiative problem, we may hope for a
rough empirical check in the photon density of the present universe.
This is a calculable quantity, though it is model dependent, and we shall
report elsewhere on the results obtained for this case
(or see~\cite{checos}).

But this is not the whole story.  There has been some previous
disagreement on the correct form of transport coefficients~\cite{and77},
especially, the bulk viscosity. The calculations are usually done in
the frame defined by Eckart and the coefficients are identified by
comparison with the general form of the stress tensor.  But there are
usually two temperatures in the expression of the stress tensor (for
matter and radiation) and the difference between them is of the same order
as the bulk viscosity. Hence the derived bulk viscosity is
dependent on which temperature is actually adopted in the stress tensor.
But with our present method, we needed to go only to first order to get
good results, so the temperature difference does not figure significantly
in the outcome.  A calculation of the entropy generation rate then leads
to an unambiguous determination of the transport coefficients.

Another way to attack this problem is to use the moment method, which
was first formally developed in kinetic theory by Grad~\cite{gra63};
in transfer theory, it may be traced back to Krook's~\cite{kro55}
formulation of Eddington's methods.  In modern times, the moment method
has been elaborated for use in radiative fluid dynamics by several
authors in attempting to deal with viscous effects, especially nonlocal
ones~\cite{and72,tho81,kat93,uda82,str97}.  However, in the context of
classical kinetic theory, the introduction of higher moments beyond those
used by Grad does not lead to rapid improvement of the macroscopic
description of rarefied gases~\cite{bgk1}.  The reason is that the
microscopic theory admits only a few slow variables, corresponding to the
lowest moments of the distribution function.

Higher moments are fast variables and, as we know from the experience of
dynamical systems theory, their introduction does not lead quickly to
improved representation of the dynamics.  Rather, it is best to attempt to
improve the description in terms of the slow variables.  Therefore, we
have pursued here an approach in which we compute only approximations for
the energy density, flux and pressure tensor that may be derived by an
expansion in the style of Thomas.  The next step in this approach would
be to seek a closure relation for the pressure tensor in terms of the
radiative variables rather than in terms of a material quantity like the
source function $S$.  We shall take this up in a later paper where we
shall see that, to improve our description in terms of the slow
variables, we should include their derivatives in the closure relation,
since the derivatives of slow variables are also slow variables.

\appendix

\section{Derivation of entropy generation rate \label{appA}}

Consider a medium made up of $N$ interacting species in a state of
near thermal equilibrium.  Let us represent the one-particle distribution
of the $i{\rm th}$ species as
\begin{equation} ^i\! f =\ ^i\! f_0 +\ ^i\! f_1
\label{app:fi}
\end{equation}
where
\begin{equation}
^i\! f_0
=exp[(-\beta_{\mu}p^{\mu}-\ ^i\!\alpha)-\ ^i\!\epsilon]
\end{equation}
is the equilibrium distribution, with $^i\! \epsilon=1,-1$ for
a Bose-Einstein or a Fermi-Dirac gas, respectively.  Also,
\begin{equation}
\beta_{\mu}=\frac{u_{\mu}}{T}
\end{equation}
and $^i\!\alpha$ is the chemical potential for the $i{\rm th}$
species, where $i=1,2,\cdots,N$. The corresponding 4-current and
stress tensor are defined as
\begin{equation}
^i\!n^{\mu}\equiv
\ ^i\!g\int\frac{p^{\mu}}{p^0}\ ^i\!fd^3p\ ,
\qquad ^iT^{\mu\nu}\equiv
\ ^i\!g\int\frac{p^{\mu}p^{\nu}}{p^0}\ ^i\!f\, d^3p
\end{equation}
while the corrections are, similarly, \begin{equation}
^i\!n^\mu_{1}\equiv\ ^i\!g
\int\frac{p^{\mu}}{p^0}\ ^i\!f_{1}d^3p
\ , \qquad ^iT_{1}^{\mu\nu}\equiv
\ ^i\!g\int\frac{p^{\mu}p^{\nu}}{p^0}\ ^i\!f_{1}d^3p
\end{equation}
where $^i\!g$ is a normalization constant depending on the nature
of the statistics of the particles.

The entropy four-vector is \begin{equation}
^i\Sigma^{\mu}= -\ ^i\!g
\int\frac{d^3p}{p^0}p^{\mu}[\ ^i\!f \ln
\ ^i\! f-\ ^i\!\epsilon
(1+\ ^i\!\epsilon)\ ^i\!f \ln(1+\ ^i\!\epsilon\ ^i\!f)].
\end{equation}
If we put equation (\ref{app:fi}) into this definition and calculate
the entropy generation rate, accurate to second order in
$^i\!f_1\, ,$ we have \begin{equation}
^i\Sigma_{\ ,\mu}^{\mu}=-\ ^i\!\alpha\  ^i\!n^{\mu}_{\ ,\mu}-
\beta_{\mu}\ ^iT_{\ ,\nu}^{\mu\nu}+\ ^i\!\epsilon
\ ^i\!\alpha_{,\mu}\ ^i\!n_{1}^{\mu}
+ \ ^i\!\epsilon\beta_{\nu,\mu}\ ^iT_{1}^{\nu\mu} \ .
\label{app:simumu}
\end{equation}
So the total entropy generation rate of the mixture is
\begin{equation}
\Sigma_{,\mu}^{\mu}=-\sum_{i}\ ^i\!\alpha \  ^i\!n^{\mu}_{\ ,\mu}-
\sum_{i} \beta_{\mu}\ ^iT_{,\nu}^{\mu\nu}+\sum_{i}\ ^i\!\epsilon
[\ ^i\!\alpha_{,\mu}\ ^i\!n_{1}^{\mu}+
\beta_{\nu,\mu}\ ^iT_{1}^{\nu\mu}]\, .
\label{app:smumu} \end{equation}

Chemical equilibrium leads to $\sum_{i} \ ^i\!\alpha
\ ^i\!n^{\mu}_{\ ,\mu}=0$ and total energy conservation results in
$\sum_{i} \beta_{\mu}\ ^iT_{,\nu}^{\mu\nu}=0$. With these two
constraints, equation(\ref{app:smumu}) reduces to:
\begin{equation}
\Sigma_{,\mu}^{\mu}=\sum_{i}\ ^i\!\epsilon
[\ ^i\!\alpha_{,\mu}\ ^i\!n_{1}^{\mu}
+\beta_{\nu,\mu}\ ^iT_{1}^{\nu\mu}] \, .  \label{app:smm} \end{equation}

With formula (\ref{app:smm}) we can evaluate the entropy generation rate
for radiating media by considering  only two species in our mixture.
The material medium, is one of these and is assumed to be in
thermal equilibrium in the absence of photons.  For this component,
$f_1=0$.  The other constituent is the radiation field, whose chemical
potential, $\alpha$, is zero. So we obtain for the entropy generation
formula of the mixture \begin{equation}
\Sigma_{,\mu}^{\mu}=\beta_{\nu,\mu}T_{1}^{\nu\mu} \; , \label{app:sm}
\end{equation}
where $T_{1}^{\nu\mu}$ is the correction to the radiative stress
tensor. With the definition for $\beta_{\nu}$, equation (\ref{app:sm})
readily reduces to the formula reported in the text except that,
in this appendix, we have not introduced a fiducial $\varepsilon.$

\section{The relation between $\dot{S}$ and $\theta$ \label{appB}}
In order to evaluate the bulk viscosity, we need to express $\dot{S}$ in
terms of $\theta$.  This can be done by analyzing the
zeroth order entropy of the entire system  --- medium plus radiation ---
and using the conservation laws.  We have the thermodynamic relation
\begin{equation}
Td\sigma=d(ev)+pdv\, ,
\label{app:aentropy}
\end{equation}
where $p$ is the total pressure, $\sigma$ is the entropy per particle,
$v$ is the volume per particle and $e$ is the total energy density.

If $T$ and $v$ are chosen to be the independent variables and $F$
is the Helmholtz energy per particle, we can write \begin{equation}
dF=-\sigma dT-pdv \; .
\label{app:helm}
\end{equation}
Since $dF$ is a perfect differential, we have the Maxwell relation
\begin{equation}
\left(\frac{\partial\sigma}{\partial v}\right)_T=\left(\frac{\partial
p}{\partial T}\right)_v \ . \end{equation}
On writing $F=ev-T\sigma$, we find $\sigma=\frac{1}{T}(ev-F)$.  If we
put this into the left of the foregoing equation and notice that from
(\ref{app:helm}), $\left(\frac{\partial F}{\partial v}\right)_T=-p$,
we get
\begin{equation}
v\left(\frac{\partial e}{\partial v}\right)_T+e+p=
T\left(\frac {\partial p}{\partial T}\right)_v \label{app:state} \ .
\end{equation}

We also have the continuity equation
\begin{equation}
\dot{n}\equiv u^{\mu}n_{,\mu}=-nu^{\mu}_{,\mu} \end{equation}
where $\dot{n}=u^{\mu}n_{,\mu}$. In terms of $v(=1/n)$, the
continuity equation can be written as
\begin{equation}
\dot{v}\equiv u^{\mu}v_{,\mu}=vu^{\mu}_{,\mu} \ .
\label{app:continuity}
\end{equation}

If we assume that the bare matter stress tensor satisfies
${\Theta}^{\mu\nu}_{(0),\nu}=0$ with
${\Theta}^{\mu\nu}_{(0)}=eu^{\mu}u^{\nu}-ph^{\mu\nu}$, we find
\begin{equation}
\dot{e}=-(e+p)\dot{v}/v \; ,
\label{app:adiabatic}
\end{equation}
where we have used (\ref{app:continuity}).  Now we are in a position
to calculate $\dot{T}$ in terms of $\theta(=u^{\mu}_{,\mu})$, starting
with the identity, \begin{equation}
e(v,T)_{,\mu}=\left(\frac{\partial e}{\partial v}\right)_T v_{,\mu}+
\left(\frac{\partial e}{\partial T}\right)_v T_{,\mu}. \end{equation}
We have  \begin{equation}
\dot{e}=\left(\frac{\partial e}{\partial v}\right)_T \dot{v}+
\left(\frac{\partial e}{\partial T}\right)_v \dot{T} \ .
\end{equation} Hence
\begin{equation}\dot{T}=\left(\frac{\partial T}{\partial
e}\right)_v\left[\dot{e}- \left(\frac{\partial e}{\partial
v}\right)_T \dot{v}\right].  \end{equation} On replacing $\dot{e}$
from (\ref{app:adiabatic}), $\left(\frac{\partial e}{\partial
v}\right)_T$ from (\ref{app:state}), and $\dot{v}$ from
(\ref{app:continuity}) in the above relation, we get
\begin{equation}
\dot{T}=u^{\mu}T_{,\mu}=-T\left(\frac{\partial p}{\partial e}\right)_v
u^{\mu}_{,\mu} \ .
\label{app:tdot}
\end{equation}
With $S=aT^4$, we find that, in the comoving frame, equation
(\ref{app:tdot}) reduces to \begin{equation}
\label{app:dotS2}
\dot{S}=-4S\left (\frac{\partial p}{\partial e}\right)_v\theta \ .
\end{equation}

\bibliographystyle{prsty}

\label{lastpage}
\end{document}